\documentclass{elsart}

\usepackage{graphics,epsfig}
\usepackage{feynarts}

\begin{document}

\begin{frontmatter}
\title{A new Supersymmetric $SU(3)_L \otimes U(1)_X$ gauge model}
\author[bogota]{Rodolfo A. Diaz}
\author[bogota]{R. Martinez}
\author[medellin]{J. Mira}
\author[bogota]{J-Alexis Rodriguez}
\address[bogota]{Departamento de F\'{\i}sica, Universidad Nacional de Colombia\\
Bogot\'a, Colombia}
\address[medellin]{Departamento de F\'{\i}sica, Universidad de Antioquia \\
Medellin, Colombia}



\begin{abstract}
We present a new  supersymmetric version of the $SU(3) \otimes U(1)$ gauge model using a more economic content of particles. The model has a smaller set of free parameters than other possibilities  considered before. The MSSM can be seen as an effective theory of this larger symmetry. We find that the upper bound of the ligthest CP-even Higgs boson can be moved up to 140 GeV.
\end{abstract}
		
\end{frontmatter}

\section{Introduction}
In recent years it has been established with great precision that interactions of the gauge bosons with the fermions are well described by the Standard Model (SM) \cite{sm}. However, some sectors of the SM have not been tested yet, this is the case of the Higgs sector, responsible for the symmetry breaking. Despite all its success, the SM still has many unanswered questions. Among the various candidates to physics beyond the SM, supersymmetric theories play a special role. Although there is not yet direct experimental evidence of Supersymmetry (SUSY), there are many theoretical arguments indicating that SUSY might be of relevance for physics beyond the SM. The most popular version, of course, is the supersymmetric version of the SM, usually called MSSM. 

Another approach to solve the fundamental problems of the SM is considering a larger symmetry which is broken to the SM symmetry using Higgs mechanisms. Among these cases 
$SU(3)_L \otimes U(1)_X$ as a local gauge theory has been studied previously by many authors who have explored different spectra of fermions and Higgs bosons \cite{31}.  There are many considerations about  this model but the most studied motivations of this large symmetry are the possibility to give mass to the neutrino sector \cite{pleitez}, anomaly  cancellations in a natural way in the 3-family version of the model, and an interpretation of the number of the fermionic families related with the anomaly cancellations \cite{su3}. 
A careful analysis of these kind of models without SUSY, has been presented recently \cite{romart}, taking into account the anomaly cancellation constraints. In fact, we supersymmetrize the version called model A in reference \cite{romart} which  has already been showm to be an anomaly free model, and  a family independent theory.

The model presented here is a supersymmetric version of the gauge symmetry $SU(3)_L \otimes U(1)_X$ but it is different from the versions considered previously \cite{ma}. The new model considered  does not introduce Higgs triplets in the spectrum to break the symmetry. Instead, they are included in the lepton superfields and the fermionic content of this new SUSY version is more economic than other ones.  As we will show, the free parameters of the model is also reduced by using a basis where only one of the vacuum expectation values (VEV) of the neutral singlets of the spectrum, generates the breaking of the larger symmetry to the SM symmetry \cite{georgi}.  Moreover, the fermionic content presented here does not have any exotic charges.

This model preserves the best features of the well motivated  $SU(3)_L \otimes U(1)_X$ symmetry and additionally when SUSY is attached 
it is  possible to shift the upper bound on the mass of the CP-even lightest Higgs boson ($h_0$). LEPII puts an experimental bound $M_h \geq 114.4$ 
GeV from direct searches of the SM Higgs boson  \cite{lep2}, but it is also known that the MSSM which is a model with two Higgs doublets imposes an upper bound on $M_h$ of about 128 GeV  \cite{bound} which up to now is consistent with the experimental bounds. In any case, the MSSM needs to find a Higgs boson around the corner, which  will be easily covered by the forthcoming LHC experiment, if it is not, the MSSM could be in trouble \cite{search,hunter}. Therefore, it is a valid motivation, to consider SUSY theories where the upper bound on $M_h$  might be moved.   

This work is organized as follows. In section 2, we present the non-SUSY version of the  $SU(3)_L \otimes U(1)_X$ model. In section 3 we discuss the SUSY version and the  spontaneous symmetry breaking mechanism,  as well as some phenomenological implications of the model. Section 4 contains our conclusions.

\section{Non-SUSY version}

We want to describe the supersymmetric version of the $SU(3)_L \otimes U(1)_X$ gauge  symmetry. But in order to be clear, first of all we present the non-SUSY version of the model. There are many possibilities for the fermionic content of the model, so we will introduce one which is economic by itself.

First of all, we present the minimal particle content. We assume that the left handed quarks and left handed leptons transform as the $\bar 3$ and $3$ representations of $SU(3)_L$ respectively. In this model the anomalies cancel individually in each family as it is done in SM. The multiplet structure for this model is
\begin{equation}
\bar Q=(\bar u, \; \bar d, \; \bar D)_L \,\; \; u^c_L \, \; \; d^c_L \, \; \; D^c_L 
\end{equation}
where they transform under the representations $(3,0)$,  $(1,-2/3)$, $(1,1/3)$, $(1, 1/3)$, respectively. For the leptonic sector, they are
\begin{equation}
L= \left( \begin{array}{c}
e^- \\ \nu_e \\ N_1^0 \end{array} \right)_L  ,\;
L_1= \left( \begin{array}{c}
E^- \\ N_2^0 \\ N_3^0 \end{array} \right)_L ,\;
L_2= \left( \begin{array}{c}
N_4^0 \\ E^+ \\ e^+ \end{array} \right)_L \; 
\label{multi}
\end{equation}
where their quantum numbers are $(3^*,-1/3)$,$(3^*,-1/3)$ and $(3^*,2/3)$ respectively. The spectrum presented in this non-SUSY model of the symmetry $SU(3)_L \otimes U(1)_X$ is  the simplest one for a single family and it is  such that $SU(3)_c \otimes SU(3)_L \otimes U(1)_X \subset E_6$ \cite{romart}. The purpose is to break down the larger symmetry in the following way:
\[
 SU(3)_L\otimes U(1)_X\longrightarrow
 SU(2)_L\otimes U(1)_Y\longrightarrow
U(1)_Q
\] 
and with this procedure  give masses to the fermion and gauge fields. To do it, we have to introduce the following set of
Higgs scalars: $\tilde L=(\bar{3},-1/3)$, $\tilde L_1(\bar{3},-1/3)$, and $\tilde L_2(\bar{3},2/3)$  which explicitly are
\begin{equation}\label{higgs}
\begin{array}{ccc}
\tilde L=\left(\begin{array}{c} \tilde l \\ \tilde N^0_1 \end{array}\right) , &
\tilde L_1=\left(\begin{array}{c} H_1 \\ \tilde N^0_3\end{array}\right) , &
\tilde L_2=\left(\begin{array}{c} H_2 \\ \tilde e^+\end{array}\right)\\
\end{array} 
\end{equation}
where $\tilde l$, $H_1$ and $H_2$ are doublet scalar fields and $\tilde N_1^0$, $\tilde N_3^0$ and $\tilde e^+$ are singlet scalar fields of $SU(2)_L$. 
We use the same letter as the fermions for the singlet scalar bosons but without the subscript that represents the quiral assigment. 

There are a total of 9 gauge bosons in the model. One gauge field
$B^\mu$ associated with $U(1)_X$,  and other 8 fields  associated
with $SU(3)_L$. The expression for the electric charge generator in  $SU(3)_L \otimes U(1)_X$ is a linear combination of the three diagonal generators of the gauge group
\begin{equation}
Q=T_{3L}+\frac{1}{\sqrt{3}}T_{8L}+XI_3
\end{equation}
where $T_{iL}=\lambda_i/2$ with $\lambda_i$ the Gell-Mann matrices and $I_3$ the unit matrix.

After breaking the symmetry, we get mass terms for the charged and the neutral gauge bosons. By diagonalizing the matrix of the neutral gauge bosons we get the physical mass eigenstates which are
defined through the mixing angle $\theta_W$ given by
$\tan \theta_W=\sqrt{3}g_1/\sqrt{3g^2+g_1^2}$. 
Also we can identify the $Y$ hypercharge associated
with the SM gauge boson as:
\begin{equation}
Y^\mu=\frac{\tan \theta_W}{\sqrt{3}}A_8^\mu+
(1-\tan \theta_W^2/3)^{1/2}B^\mu.
\end{equation}

In the SM the coupling constant $g'$ associated with the hypercharge $U(1)_Y$,
can be given by $\tan \theta_W=g'/g$ where $g$ is the coupling constant of $SU(2)_L$ which in turn can be taken equal to the $SU(3)_L$ coupling constant. Using the $\tan \theta_W$ given by the diagonalization of the neutral gauge boson matrix, we obtain the matching condition
\begin{equation}\label{gut}
\frac{1}{g'^2}=\frac{1}{g_1^2}+\frac{1}{3g^2}\;,
\label{matching}
\end{equation}
where $g_1$ is the coupling constant associated to $U(1)_X$. We shall use this relation to write $g_1$ as a function of $g'$ in order to find the potential of the $SU(3)\otimes U(1)_X$ SUSY model at
low energies and compare it with the MSSM one. In particular, we will show that it reduces to the MSSM in this limit.

 
\section{SUSY version}

In the SUSY version the above content of fermions should be  written in terms of chiral superfields, and the gauge fields will be in vector supermultiplets as it is customary in SUSY theories. One more ingredient may be taken into account due to the possibility of having  terms which  contribute to baryon number violation and fast proton decay. It is a discrete symmetry $Z_2$ which avoids these kind of terms, explicitly it reads
\begin{eqnarray}
(\hat Q, \hat u, \hat L, \hat L_1) &\rightarrow& (-\hat Q, - \hat u, -\hat L, -\hat L_1) \nonumber \\
(\hat L_2, \hat d ,\hat D) &\rightarrow& (\hat L_2, \hat d ,\hat D).
\end{eqnarray}
Then, we build up the superpotential
\begin{eqnarray}
W&=&h_e \epsilon_{abc}\hat L^a \hat L_1^b \hat L_2^c + h^u Q L_2 U+ h^d Q L_1 d + h^D Q L_1 D \nonumber \\
&+& h^1 Q L_1 D + h^2 Q L d  ,
\end{eqnarray}
 which is  invariant under SUSY,  $SU(3)\otimes U(1)_X$  and  $Z_2$ symmetries.
In our analysis the first term is the most relevant, because we shall deal with the scalar sector mainly and it is going to introduce new physics at low energies. We can note that the scalar sector of the leptonic superfields can be used as Higgs bosons adequately, see equation (\ref{higgs}). This fact is attractive  because it makes the model economic in its particle content. Therefore, this SUSY version does not require additional chiral supermultiplets which include the Higgs sector in their scalar components. Instead, we have the Higgs fields in the scalar components of our lepton multiplets (eq. \ref{multi}) because they have the right quantum numbers that we need for the Higgs bosons, eq. (\ref{higgs}). Also with this arrangement of fermions the SUSY model is triangle anomaly free.

In general, it is possible that the neutral scalar particles $\tilde \nu$, $\tilde N_1^0$, $\tilde N_2^0$,  $\tilde N_3^0$, and  $\tilde N_4^0$ can get VEV's different from zero. But, in order to break down the larger symmetry   $SU(3)_L \otimes U(1)_X$ we will consider as a first step that only  $\tilde N_{1,3}^0$  acquire a nonzero VEV, and later on, the $H_{1,2}$ fields break down the SM symmetry. We should mention that it is possible to reduce the free parameters of the theory by choosing a convenient basis.  In the first step, we will choose $\langle \tilde N_3 \rangle =0$ \cite{georgi}.

Once we have the superpotential $W$, the theory is defined and we can get the Yukawa interactions and the scalar potential. We will concentrate our attention on  the scalar potential, which is given by
\begin{equation}
V=\vert \frac {\partial W}{\partial A_i}\vert^2 + \frac 1 2 D^a D^a + \frac 1 2 D' D'
\end{equation}
where
\begin{eqnarray}
D^a &=& g A_i^{\dagger} \frac{\lambda_{ij}}{2} A_j \; \nonumber ,\\
D' &=& g_1 A_i^{\dagger} X(A_i) A_i  
\end{eqnarray}
and $A_i$ are the scalar components of the chiral supermultiplets, equation (\ref{multi}). The prescription yields
\begin{eqnarray}
V&=& \frac{g^2}{6}\left[ (L^\dagger L)^2 + (L_1^\dagger L_1)^2 +(L_2^\dagger L_2)^2+ 3 (L^\dagger L_1)^2 - (L^\dagger L)(L_1^\dagger L_1) \right. \nonumber \\
&+&  \left. 3 (L^\dagger L_2)^2 - (L^\dagger L)(L_2^\dagger L_2)+ 3(L_1^\dagger L_2)^2-(L_1^\dagger L_1)(L_2^\dagger L_2)\right] \nonumber \\
&+&\frac{g_1^2}{18} \left[ (L^\dagger L)^2 + (L_1^\dagger L_1)^2 +4 (L_2^\dagger L_2)^2 + 2 (L^\dagger L)(L_1^\dagger L_1)-4(L^\dagger L)(L_2^\dagger L_2)  
\right. \nonumber \\
&-& \left. 4 (L_1^\dagger L_1)(L_2^\dagger L_2) \right] \nonumber \\
&+& h_e^2 \left[(L_1^\dagger L_1)(L_2^\dagger L_2)-(L_1^\dagger L_2)(L_2^\dagger L_1)+(L^\dagger L)(L_2^\dagger L_2)-(L^\dagger L_2)(L_2^\dagger L) \right. \nonumber \\
&+& \left.(L^\dagger L)(L_1^\dagger L_1)-(L^\dagger L_1)(L_1^\dagger L) \right] \; ,
\end{eqnarray}
and the soft terms that  only affects the scalar potential considered are
\begin{eqnarray}
V_{soft}&=&m_L^2 \tilde L^\dagger \tilde L + m_{L_2}^2 \tilde L_2^\dagger \tilde L_2 + m_{L_1}^2 \tilde L_1^\dagger \tilde L_1 +m_{L L_1} (\tilde L^\dagger \tilde L_1+ h.c)\nonumber \\
&+& h' (\epsilon_{abc}\tilde L^a \tilde L_1^b \tilde L_2^c+ h.c.).
\end{eqnarray}

Now, we are ready to break down the symmetry  $SU(3)_L \otimes U(1)_X$ to the SM symmetry  $SU(2)_L \otimes U(1)_Y$. Thus the VEV's of $< \tilde N_1^0>=u$ and $< \tilde N_3^0>=u'$, will make the job. But it is possible to choose one of them to be zero, e.g. $u'=0$ \cite{georgi}, and the would-be Goldstone bosons of the symmetry breaking $SU(3)_L \otimes U(1)_X \big/ SU(2)_L\otimes U(1)_Y$  become degrees of freedom of the field $\tilde L$. Further, if we choose our basis in the mentioned way, we decouple the fields in $\tilde L$ and $\Re(\tilde N_3^0)$ from the electroweak scale where the remnant symmetry is $SU(2)_L \otimes U(1)_Y$. 

In order to get the reduced Higgs potential we introduce the following definitions
\begin{equation} 
H_1=\left(\begin{array}{c} -\tilde E^- \\ \tilde N_2^0 \end{array}\right) \; ,\; 
H_2=\left(\begin{array}{c} -\tilde N_4^0 \\ -\tilde E^+\end{array}\right) \; , \; 
\tilde l=\left(\begin{array}{c} -\tilde e^- \\ \tilde\nu\end{array}\right)
\end{equation}
and therefore the scalar components of our superfields are precisely written as equation (\ref{higgs}).  We should note that the arrays $\tilde l$ and $H_{1(2)}$ transform under the conjugate representation $2^*$ of $SU(2)_L$ meanwhile the fields $\tilde N_{1,3}^0$ are singlets. 
\begin{figure}[htbp]
 {\hbox{
    \includegraphics[width=7cm]{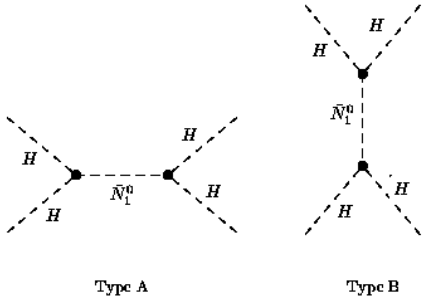} }}
 \caption{\it Diagrams type A and B which contribute to the effective couplings. External legs are  bosons $H_{1,2}$  \it and the exchanged boson is the heavy $\tilde N_1^0$ \it field. }   
\end{figure}
With the above definitions we see that the parts of the potential which contain $H_1$, $H_2$ and $\tilde N_1$ are
\begin{eqnarray}
V'&=&\frac{g^2}{6}\left[ (\tilde N_1^2)^2 + (H_1^\dagger H_1)^2 + (H_2^\dagger H_2)^2  + 3(\epsilon_{ij}H_1^i H_2^j)^2 \right. \nonumber \\
&-& \left.\tilde N_1^2(H_2^\dagger H_2)-(H_1^\dagger H_1)(H_2^\dagger H_2) +\tilde N_1^2(H_1^\dagger H_1)\right] \nonumber \\
&+& \frac{g_1^2}{18}\left[ 4(H_2^\dagger H_2)^2+(\tilde N_1^2)^2+(H_1^\dagger H_1)^2+2 \tilde N_1^2(H_1^\dagger H_1) \right. \\
&-& \left. 4 \tilde N_1^2(H_2^\dagger H_2)-4(H_1^\dagger H_1)(H_2^\dagger H_2)\right] \nonumber \\
&+& h_e^2 \left[  (H_1^\dagger H_1)(H_2^\dagger H_2)-
 (\epsilon_{ij}H_1^i H_2^j)^2+ \tilde N_1^2(H_2^\dagger H_2)+\tilde N_1^2(H_1^\dagger H_1)\right]. \nonumber
\end{eqnarray}
The minimum conditions for the potential with the VEV $< \tilde N_1^0>=u$ and $< \tilde N_3^0>=u'$ when $u'$ goes to zero are satisfied if
\begin{eqnarray}
m_{L \phi}^2&=& 0 \nonumber \\
m_L^2&=& -\left( \frac{{{g_1}}^2}{9} + \frac{{{g}}^2}{3} \right)\,u^2 
\end{eqnarray}
and, therefore the mass of the field $\tilde N_1^0$ is given by $m_N^2= 4 \left( \frac{{{g_1}}^2}{9} + \frac{{{g}}^2}{3} \right)\,u^2$.

In the  MSSM two scalar doublets appear, it is because their fermionic partners are necessary to cancel the axial-vector triangle anomalies. The requirement of SUSY also constrains the parameters of the Higgs potential. Therefore the Higgs potential of the MSSM  can be seen as a special case of the more general 2HDM potential structure. This result in constraints among the $\lambda_i$'s  of the general 2HDM potential \cite{ma,hunter}.

As it has been already emphasized \cite{ma}, in the MSSM the quartic scalar couplings of the Higgs potential are completely determined in terms of the two gauge couplings, but it is not the case if the symmetry $SU(2)_L \otimes U(1)_Y$  is a remnant of a larger symmetry which is broken at a higher mass scale together with the SUSY. The structure of the Higgs potential is then determined by the scalar particle content needed to produce the spontaneous symmetry breaking. In this way, the reduced Higgs potential would be a 2HDM-like potential, but its quartic couplings would not be those of the MSSM. Instead, they will be related to the gauge couplings of the larger theory and to the couplings appearing in its superpotential. Analysis of supersymmetric theories  in this context  have been given in the literature \cite{ma,others}. In particular, it has been studied widely for different versions of the left-right model and a specific SUSY version of the $SU(3)_L \otimes U(1)_X$ where exotic charged particles of electric charges $(-4/3, 5/3)$ appear.

Following this idea with the reduced Higgs potential presented in the previous paragraphs, we can obtain the effective quartic scalar couplings $\lambda_i$ of the most general 2HDM potential. Since there are cubic interactions in $V'$ involving $H_{1,2}$ and $\tilde N_1^0$, it generates two types of Feynman diagrams which contribute to the quartic couplings (figure 1). The Feynman rules from the potential for these couplings are

\unitlength=0.8bp%

\begin{feynartspicture}(332,332)(2,2)

\FADiagram{}
\FAProp(0.,15.)(9.,10.)(0.,){/ScalarDash}{0}
\FALabel(4.6,15.8436)[tr]{$H_2$}
\FAProp(0.,5.)(9.,10.)(0.,){/ScalarDash}{0}
\FALabel(4.65134,6.8436)[tl]{$H_2$}
\FAProp(20.,10.)(9.,10.)(0.,){/ScalarDash}{0}
\FALabel(31.5,9.12)[b]{$ i \left(\frac{-4\,{{g_1}}^2}{9} - \frac{{{g}}^2}{3} + 2 \,h_e^2\right) u$}
\FALabel(14.5,10.82)[b]{$\tilde N_1^0$}
\FAVert(9.,10.){0}
\FADiagram{}
\FADiagram{}
\FAProp(0.,15.)(9.,10.)(0.,){/ScalarDash}{0}
\FALabel(4.6,15.8436)[tr]{$H_1$}
\FAProp(0.,5.)(9.,10.)(0.,){/ScalarDash}{0}
\FALabel(4.65134,6.8436)[tl]{$H_1$}
\FAProp(20.,10.)(9.,10.)(0.,){/ScalarDash}{0}
\FALabel(31.5,9.12)[b]{$ i \left(\frac{2{{g_1}}^2}{9} - \frac{{{g}}^2}{3} + 2 \,h_e^2\right) u$}
\FALabel(14.5,10.82)[b]{$\tilde N_1^0$}
\FAVert(9.,10.){0}







\end{feynartspicture}

and using them we obtain the effective couplings, taking into account that the diagrams presented in figure 1 contribute to $\lambda_{1,2,3}$; thus they are given by
\begin{eqnarray}
\frac {\lambda_1}{2} &=& \frac{g_1^2}{18}+ \frac{g^2}{6}-\frac 3 8 
\left((\frac{2 g_1^2}{9}-\frac{g^2}{3})^2+4 h_e^2(\frac{2 g_1^2}{9}-\frac{g^2}{3}) + 4h_e^4 \right) G^{-1} \; \; ,\nonumber \\
\frac {\lambda_2}{2} &=& \frac{2 g_1^2}{9}+ \frac{g^2}{6}-\frac 3 8 
\left((\frac{4 g_1^2}{9}+\frac{g^2}{3})^2- 4 h_e^2(\frac{4 g_1^2}{9}+\frac{g^2}{3}) + 4h_e^4 \right) G^{-1} \; \; ,\nonumber \\ 
\lambda_3 &=& -\frac{2 g_1^2}{9}- \frac{g^2}{6}+\frac 3 4 
\left((\frac{4 g_1^2}{9}+\frac{g^2}{3})(\frac{2 g_1^2}{9}-\frac{g^2}{3})\right. \nonumber \\ 
&+&\left. 4 h_e^2(\frac{g_1^2}{9}+\frac{g^2}{3}) - 4h_e^4 \right) G^{-1} \; \; ,\nonumber \\ 
\lambda_4&=&\frac{g^2}{2}-h_e^2 \; \; \; , \; \; \lambda_5=0 
\end{eqnarray}
\begin{figure}[htbp]
 {\hbox{
    \includegraphics[width=10cm]{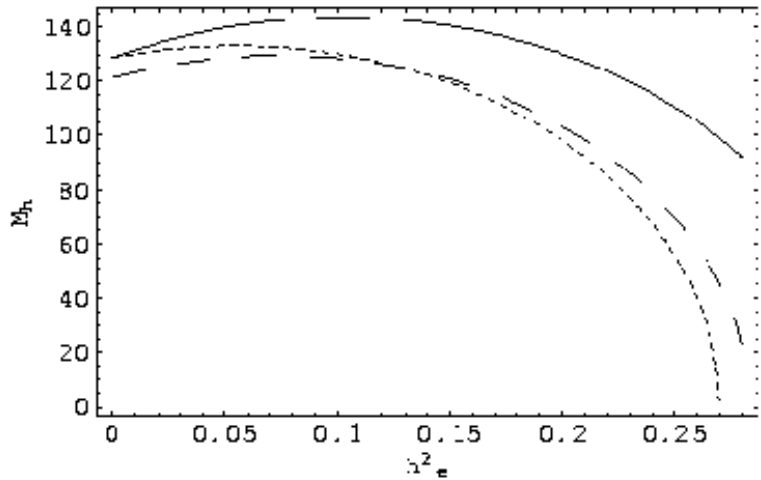} }}
 \caption{\it The upper bound on the lightest CP-even Higgs boson of the model as a function of the parameter $h_e^2$ from the superpotential. The  solid, dashed and dotted lines correspond to $\cos^2 \beta =\{0\;,\;0.6\;,\;1\}$ respectively. }   
\end{figure}
where $G= \left(\frac{g_1^2}{3}+g^2 \right)$.  We want to remark that this SUSY model has the MSSM as an effective theory when the new physics is not longer there, $h_e=0$, and the coupling constants are running down to the electroweak scale.  At this point we use the approach where the $SU(2)_L$ coupling behaves like $g$, and $g_1$ is the combination given by (\ref{matching}). In the limit $h_e^2=0$, we obtain
\[
\lambda_{1,2} = \frac {g^2(4g_1^2+3g^2)}{4(g_1^2+3g^2)} \; \; , \; \;
\lambda_3 = - \frac {g^2(4g_1^2+3g^2)}{4(g_1^2+3g^2)} \, \, , \,\, 
\lambda_4=\frac{g^2}{2} \nonumber
\]
and, if we assume the matching condition from equation (\ref{gut}), we reduce the effective couplings to those appearing in the MSSM, as expected,
\[
\lambda_1=\lambda_2=\frac 14 (g^2+g'^2) \; \; ,\;\;
\lambda_3= - \frac 14 (g^2+g'^2) \, \, , \, \, \lambda_4 =\frac{g^2}{2}. \nonumber
\]

When we have the reduced Higgs potential, we should ask for the stability conditions. These  conditions are well known \cite{hunter} and they give us a constrain for the coupling $h_e$ which is a coupling in the superpotential of the larger symmetry and a new free parameter. The general requirement for $V$ to be bounded from below leads to the allowed region, $0 \leq h_e^2 \leq 0.28$.

On the other hand, for $\lambda_5=0$ in the potential, there is a general formula to obtain the upper bound on $M_h$ in the framework of a general Two Higgs Doublet Model \cite{hunter,ma}. This formula is given in terms of the $\lambda_i$ parameters and we use it along with equation (18) in order to make plots in the $M_h$-$h_e^2$ plane. Figure 2 shows the plane $M_h$ versus $h_e^2$, for different values of $\cos^2 \beta$, where $\beta$ is the CP-odd mixing angle. It is obvious that we can move the lower bound predicted by MSSM of about 128 GeV according to the values  of the parameters involved in this model. In particular we get the upper bound of MSSM in the limit $h_e=0$. Further, we note from figure 2  that we can shift the upper bound up to a value of around 140 GeV for $h_e^2=0.1$ and $\cos \beta=0$. The upper bound on $M_h$ is consistent with the experimental bound from LEPII.

\section{CONCLUSIONS}
We have presented a new supersymmetric version of the gauge symmetry $SU(3)_L \otimes U(1)_X$ where the Higgs bosons correspond to the sleptons, and it is triangle anomaly free. The model has a number of free parameters which is smaller than other ones in the literature. We have also shown that using the limit when the parameter $h_e=0$ and  the matching condition (equation (6)), we obtain the SUSY constraints 
for the Higgs potential as in the MSSM. Therefore, if we analyze the upper bound for the mass of the lightest CP-even Higgs boson in this limit, we find the 
same bound of around 128 GeV for the MSSM. However, since in general $h_e \not = 0$, such upper bound can be moved up to around 140 GeV, see figure 2. This fact can be an interesting alternative to take into account in the search for the Higgs boson mass.

We acknowledge to D. Restrepo, W. Ponce and L. Sanchez for useful disscussions.This work was supported by COLCIENCIAS, DIB and Banco de la Republica.

\end{document}